\documentclass{article}
\usepackage{spconf,amsmath,graphicx}
\usepackage{array}
\usepackage{booktabs}
\usepackage[caption=false,font=normalsize,labelfont=sf,textfont=sf]{subfig}
\usepackage{textcomp}
\usepackage{stfloats}
\usepackage{url}
\usepackage{verbatim}
\usepackage{graphicx}
\usepackage{cite}

\usepackage[nolist,nohyperlinks]{acronym}
\usepackage{bm}
\usepackage{comment}
\usepackage{adjustbox}
\usepackage{multirow, multicol}
\usepackage{orcidlink}
\usepackage{acronym}
\usepackage{amsfonts}
\usepackage{xcolor}
\usepackage{hyperref}
\usepackage{colortbl}

\usepackage{cleveref}
\begin{acronym}
\acro{AE}{autoencoder}
\acro{ASV}{automatic speaker verification}
\acro{BCE}{binary cross-entropy}
\acro{CM}{countermeasure}
\acro{CNN}{convolutional neural network}
\acro{COTS}{commercial-off-the-shelf}
\acro{CQCC}{constant-Q cepstral coefficient}
\acro{CRNN}{convolutional recurrent neural network}
\acro{DNN}{deep neural network}
\acro{DF}{deepfake}
\acro{EER}{equal error rate}
\acro{ELU}{exponential linear unit}
\acro{FCN}{fully convolutional network}
\acro{GMM}{Gaussian mixture model}
\acro{GRU}{gated recurrent unit}
\acro{IoT}{Internet of things}
\acro{IPD}{interchannel phase difference}
\acro{IR}{impulse response}
\acro{LA}{Logical access}
\acro{LSTM}{long short-term memory}
\acro{MVDR}{minimum variance distortionless response}
\acro{MWF}{multichannel Wiener filtering}
\acro{ReMASC}{Realistic Replay Attack Microphone Array Speech}
\acro{RIR}{room impulse response}
\acro{RNN}{recurrent neural network}
\acro{ROC}{receiving operating characteristic}
\acro{TTS}{text-to-speech}
\acro{PA}{physical access}
\acro{PFA}{probability of false alarm}
\acro{SNR}{signal-to-noise ration}
\acro{SOTA}{state-of-the-art}
\acro{STFT}{short-time Fourier transform}
\acro{VA}{voice assistant}
\acro{VC}{voice conversion}
\end{acronym}

\definecolor{Gray}{gray}{0.85}
\definecolor{red_cool}{rgb}{0.5, 0.0, 0.0}

\definecolor{light_red}{HTML}{FCBABA}
\definecolor{light_green}{HTML}{E4F2D5}
\definecolor{light_yellow}{HTML}{FCF6E1}

\title{Acoustic Simulation Framework for Multi-channel Replay Speech Detection}
\name{Michael~Neri~\orcidlink{0000-0002-6212-9139},      
      Tuomas~Virtanen~\orcidlink{0000-0002-4604-9729}}
\address{\textit{Faculty of Information Technology and Communication Sciences, Tampere University, Tampere, Finland}\\
         \textit{\{michael.neri, tuomas.virtanen\}@tuni.fi}\\
}

\begin{document}
\ninept
\maketitle
\begin{abstract}
Replay speech attacks pose a significant threat to voice-controlled systems, especially in smart environments where voice assistants are widely deployed. While multi-channel audio offers spatial cues that can enhance replay detection robustness, existing datasets and methods predominantly rely on single-channel recordings. Moreover, previous studies highlighted that generalization of this attack to new environments is challenging, requiring new methods for generating data encompassing various acoustic conditions. Hence, in this work we introduce an acoustic simulation framework designed to simulate multi-channel replay speech configurations using publicly available resources. Using the framework, we train the state-of-the-art multi-channel replay detector M-ALRAD and evaluate its generalisation on the ReMASC real-recording corpus without any real training data. To improve the exploitation of spatial information, we extend M-ALRAD with inter-channel phase difference features computed for adjacent microphone pairs, augmenting the beamformed representation with directional cues. Synthetic datasets are available at \href{https://github.com/michaelneri/synthetic-ReMASC}{\textit{https://github.com/michaelneri/synthetic-ReMASC}}.
\end{abstract}
\begin{keywords}
Replay attack, physical access, spatial audio, voice spoofing, room acoustic simulation
\end{keywords}
\section{Introduction}
\label{sec:intro}

The use of \acp{VA} has become increasingly prevalent in human-machine interaction, using voice as a biometric identifier~\cite{Huang_IoTJ_2022}. They enable users to control smart devices via \ac{IoT} networks and exchange sensitive information online. This growing reliance has made audio-based systems a target for adversarial attacks exploiting vulnerabilities in speech recordings and attacking the \ac{ASV}~\cite{Liu_TASLP_2023}. \ac{LA} attacks manipulate speech using \ac{TTS} and \ac{VC} techniques to mimic a speaker, while \ac{DF} methods further obscure artifacts through compression and quantization. In contrast, \ac{PA} attacks deceive the \ac{ASV} system at the microphone level~\cite{Kinnunen_ICASSP_2017, Sanchez_TIFS_2026}, including \textit{impersonation attacks}~\cite{Huang_IoTJ_2022} and \textit{replay attacks}~\cite{Gong_SPL_2020}, where recorded speech is played back to gain unauthorized access.

\begin{figure}[ht!]
    \centering
    \includegraphics[width=\linewidth]{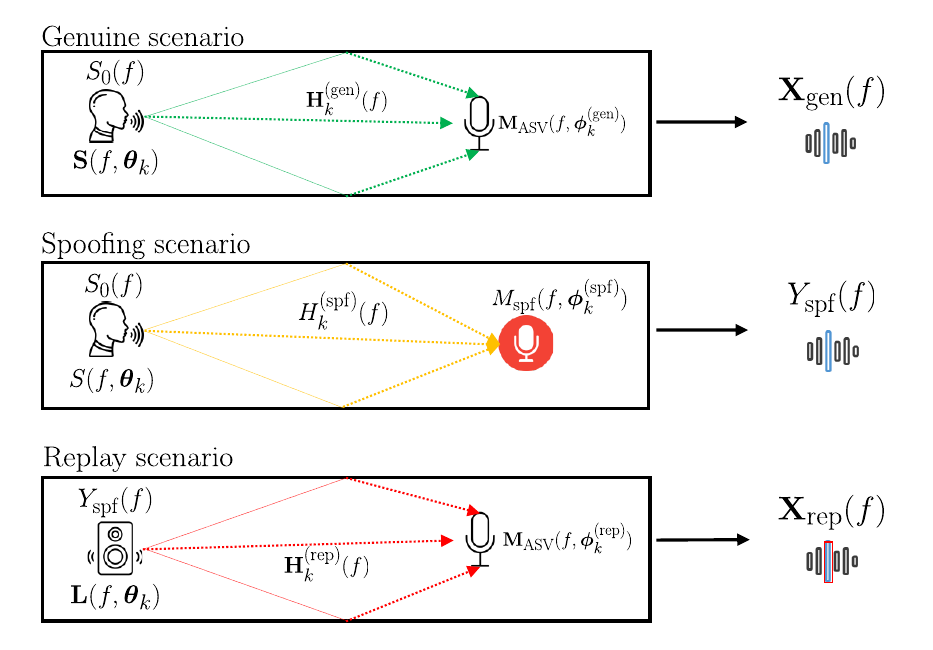}
    \caption{Overview of the acoustic simulation framework for replay speech detection.}
    \label{fig:overview}
\end{figure}

Previous works to address replay attacks have primarily relied on single-channel datasets, such as RedDots~\cite{Kinnunen_ICASSP_2017} and the ASVSpoof \ac{PA} series~\cite{kinnunen2017asvspoof, todisco2019asvspoof, Liu_TASLP_2023}. These resources have supported the development of replay detection models based on \ac{DNN} architectures~\cite{Luo_ICASSP_2021, Xue_LSP_2024} and hand-crafted time-frequency features~\cite{boyd2023voice, Xu_TASLP_2023}.  However, microphone arrays are commonly integrated into \ac{ASV} systems to improve tasks such as speech enhancement and separation by leveraging spatial information~\cite{omologo2001speech}. Beyond audio quality, multi-channel data provides key advantages for replay attack detection. Specifically, (i) spatial cues in multi-channel recordings support more robust detection~\cite{Gong_Interspeech_2019, Gong_SPL_2020}, and (ii) these spatial features are difficult for attackers to replicate, unlike temporal or spectral cues in single-channel data~\cite{zhang2016voicelive}. Nevertheless, most existing replay detectors rely on single-channel input and fail to exploit spatial information~\cite{Gong_SPL_2020, Neri_OJSP_2025, Neri_EUSIPCO_2025}. This is aggravated by the fact that the \ac{ReMASC} dataset~\cite{Gong_Interspeech_2019} is the only replay speech dataset that encompasses real recordings from four microphone arrays and four environments. In fact, hardware costs, the need for synchronization and calibration, the need for a lot of storage, and the difficulty of reproducing consistent spatial setups across a variety of environments make it difficult to collect multi-channel and spatial audio datasets. Furthermore, generalization is limited to unseen conditions due to the labor-intensive nature of obtaining precise ground truth spatial labels, maintaining privacy, and achieving dataset diversity. Hence, it is challenging to evaluate the impact of each single component in the replay speech task. Moreover, several prior studies on replay speech detection (both single- and multi-channel setups)~\cite{Sanchez_TIFS_2026, Neri_OJSP_2025} highlighted that generalizing to unseen environments is one of the most difficult challenges. 

To address the aforementioned problems, the contributions of this work are as follows: (i) we propose an acoustic simulation framework that generates multi-channel replay speech recordings, enabling controlled experiments on the factors affecting replay speech detection, (ii) we extend M-ALRAD~\cite{Neri_OJSP_2025} with inter-channel phase difference features computed for adjacent microphone pairs, augmenting the beamformed representation with spatial cues that generalise more reliably from synthetic to real recordings than single-channel spectral features, and (iii) we conduct a environment-independent evaluation on the ReMASC corpus, training exclusively on synthetic data, and perform an ablation study over the beamformer, per-channel log-power, and \ac{IPD} branches, analysing which features are more generalisable to unseen real environments.

\section{Replay speech simulation}

In a typical replay attack condition, a talker’s speech is captured both by the verification system’s microphone array and by an attacker’s recording device, either concurrently or independently. The attacker later replays the captured signal through a loudspeaker (or other playback device), and the detector has to determine whether the received signal is live or replayed~\cite{Sanchez_TIFS_2026}.

We simulate the replay attack defining three scenarios, using the frequency-domain notation~\cite{benesty2008microphone} in Table~\ref{tab:notation}. In the \textit{genuine scenario}, the \ac{ASV} system records the speech from the talker, which can be modeled as a sum of different propagation paths $K_{\mathrm{gen}}$. In particular, the source speech spectrum $S_0(f)$ is filtered by the frequency-dependent directional responses $\mathbf{S}(f,\boldsymbol{\theta}_k)$ of the talker along each acoustic propagation path $k$ in the room having path-wise azimuth and elevation angles $\boldsymbol{\theta}_k=[\theta_k^{\mathrm{azimuth}}, \theta_k^{\mathrm{elevation}}]$ at which the path leaves the talker. Specifically, $\mathbf{S}(f,\boldsymbol{\theta}_k)$ is a vector which defines directional responses on the path to each detection microphone. Then, the source speech spectrum is filtered by the path-dependent acoustic transfer function $\mathbf{H}^{\mathrm{(gen)}}_k(f)$ which consists of a vector of transfer functions, where each entry corresponds to one path to each detection microphones. Similarly, we define the directional transfer function of each \ac{ASV} microphone as the vector $\mathbf{M}_{\mathrm{ASV}}(f,\boldsymbol{\phi}^{\mathrm{(gen)}}_k)$ with arrival angles $\boldsymbol{\phi}_k^{\mathrm{(gen)}}$. The genuine multi-channel speech recorded from the \ac{ASV} array $\mathbf{X}_{\mathrm{gen}}(f)$ can be expressed as

\begin{align}
\mathbf{X}_{\mathrm{gen}}(f)={}&
S_0(f)\sum_{k}
\mathbf{S}\bigl(f,\boldsymbol{\theta}_k\bigr)\odot
\mathbf{H}^{\mathrm{(gen)}}_k(f)\odot
\mathbf{M}_{\mathrm{ASV}}\bigl(f,\boldsymbol{\phi}^{\mathrm{(gen)}}_k\bigr)\nonumber\\
&+\mathbf{N}(f),
\label{eq:genuine}
\end{align}
where $\odot$ denotes element-wise multiplication of vectors and $\mathbf{N}(f)$ is the multi-channel background noise recorded from the \ac{ASV} array.

In the \textit{spoofing scenario} of the attack, an adversary places a single-channel \textit{spoofing microphone} close to the talker which records the  source signal through its own set of $K_{\mathrm{spf}}$ propagation paths. Each path applies its own direction-dependent transfer function $S(f,\boldsymbol{\theta}_k)$ to model the source directivity, path-dependent transfer function $H_k^{\mathrm{(spf)}}(f)$ to model room acoustics, and directional spoof-microphone response $M_{\mathrm{spf}}(f,\boldsymbol{\phi}_k^{\mathrm{(spf)}})$, resulting in the single-channel spoofed signal with noise $N(f)$

\begin{equation}
Y_{\mathrm{spf}}(f)
=
S_0(f)\,
\sum_{k}
S\!\bigl(f,\boldsymbol{\theta}_k\bigr)\,H_k^{\mathrm{(spf)}}(f)\,
M_{\mathrm{spf}}\!\bigl(f, \boldsymbol{\phi}^{\mathrm{(spf)}}_k \bigr)+ N(f).
\label{eq:spoof}
\end{equation}

Finally, in the \textit{replay scenario}, the attacker replays $Y_{\mathrm{spf}}(f)$ through a loudspeaker with directional radiation pattern $\mathbf{L}(f,\boldsymbol{\theta}_k)$ applied to all propagation paths $K_{\mathrm{rep}}$ in the room having path-wise azimuth and elevation angles $\boldsymbol{\theta}_k=[\theta_k^{\mathrm{azimuth}}, \theta_k^{\mathrm{elevation}}] $ at which each path leaves the source. Similarly with the speech signal, $\mathbf{L}(f,\boldsymbol{\theta}_k)$ is a vector encompassing directional responses for each detection microphone. Then, the replayed signal propagates along $K_{\mathrm{rep}}$ replay paths, each with its own acoustic transfer function of the room $\mathbf{H}^{\mathrm{(rep)}}_k(f)$ and directional microphone array response $\mathbf{M}_{\mathrm{ASV}}(f,\boldsymbol{\phi}_k^{\mathrm{(rep)}})$. Again, propagation and microphone responses are combined element-wise across channels before summation, leading to the final multichannel replay signal

\begin{align}
\mathbf{X}_{\mathrm{rep}}(f)={}&
Y_{\mathrm{spf}}(f)\sum_{k}
\mathbf{L}\bigl(f,\boldsymbol{\theta}_k\bigr)\odot
\mathbf{H}_k^{\mathrm{(rep)}}(f)\odot
\mathbf{M}_{\mathrm{ASV}}\bigl(f,\boldsymbol{\phi}_k^{\mathrm{(rep)}}\bigr)\nonumber\\
&+\mathbf{N}(f).
\label{eq:replay}
\end{align}

Because of the replay chain, which results in temporal smearing or \textit{double} reverberation, replay signals are different from real ones and produce quantifiable artifacts~\cite{Gaubitch_ICASSP_2024}. Therefore, in order to generalize robustly, detection systems rely on features that are sensitive to re-recording artifacts, device/room characteristics, and directionalities.

\begin{table}[ht!]
\vspace{-6pt}
\centering
\caption{Notation used for the description of the replay speech attack.}
\adjustbox{max width=0.5\textwidth}{%
\begin{tabular}{cc}
\toprule
\textbf{Symbol} & \textbf{Meaning, Typical Values / Sets, and Field} \\
\midrule
$K_{\mathrm{gen}}, K_{\mathrm{spf}}, K_{\mathrm{rep}}$ & $\mathbb{Z}_{\ge 1}$: Number of propagation paths in each scenario. \\
$C$ & $\mathbb{Z}_{\ge 1}$: Number of ASV array microphones. \\
\midrule
$S_0(f)$ &  $\mathbb{C}:$ Source speech spectrum. \\
$N(f)$ &  $\mathbb{C}:$ Single-channel diffuse noise spectrum. \\
$\mathbf{N}(f)$ &  $\mathbb{C}^{C\times 1}:$ Multi-channel diffuse noise spectrum. \\
$\boldsymbol{\theta}_k$ & $\mathbb{R}^2$: Source or loudspeaker radiation direction for path $k$. \\
$\boldsymbol{\phi}_k$ & $\mathbb{R}^2$: Arrival direction to the microphones for path $k$. \\
$\mathbf{S}(f,\boldsymbol{\theta}_k)$ & $\mathbb{C}^{C \times 1}$ :Talker directivity for path $k$, $k = 1,\dots,K_{\mathrm{gen}}$ or $K_{\mathrm{spf}}$. \\
$\mathbf{H}_k^{\mathrm{(gen)}}(f)$ & $\mathbb{C}^{C \times 1}$: Acoustic transfer function of path $k$ to \ac{ASV} array. \\
$\mathbf{M}_{\mathrm{ASV}}(f,\boldsymbol{\phi}_k^{\mathrm{(gen)}})$ &  $\mathbb{C}^{C \times 1}$: ASV array directional response for path $k$. \\
\midrule
$H_k^{\mathrm{(spf)}}(f)$ & $\mathbb{C}$: Acoustic transfer function of path $k$ to spoofing microphone. \\
$M_{\mathrm{spf}}(f,\boldsymbol{\phi}_k^{\mathrm{(spf)}})$ & $\mathbb{C}$: Spoofing microphone directional response for path $k$. \\
\midrule
$\mathbf{L}(f,\boldsymbol{\theta}_k)$ & $\mathbb{C}^{C \times 1}$: Loudspeaker directivity for replay path $k$, $k = 1,\dots,K_{\mathrm{rep}}$. \\
$\mathbf{H}_k^{\mathrm{(rep)}}(f)$ & $\mathbb{C}^{C \times 1}$: Acoustic transfer function of path $k$ from loudspeaker to \ac{ASV} array. \\
$\mathbf{M}_{\mathrm{ASV}}(f,\boldsymbol{\phi}_k^{\mathrm{(rep)}})$ & $\mathbb{C}^{C \times 1}$: \ac{ASV} array directional response for replay path $k$. \\
\midrule
$\mathbf{X}_{\mathrm{gen}}(f)$ & $\mathbb{C}^{C \times 1}$: Multichannel signal recorded by \ac{ASV} array during genuine scenario. \\
$Y_{\mathrm{spf}}(f)$ & $\mathbb{C}$: Single-channel recording by spoofing microphone. \\
$\mathbf{X}_{\mathrm{rep}}(f)$ & $\mathbb{C}^{C \times 1}$: Multichannel replayed signal captured by \ac{ASV} array. \\
\bottomrule
\end{tabular}
}
\label{tab:notation}
\vspace{-5pt}
\end{table}

\section{Materials}\label{sec:material}
In this section we describe the used materials for the construction of the replay speech synthetic dataset. The overall generation framework is depicted in Figure~\ref{fig:overview}.

\subsection{Speech and Loudspeaker Directivities}
The following open source datasets have been employed in the simulation framework: 

(i) EARS~\cite{Richter_Interspeech_2024} dataset is a collection of anechoic speech recordings featuring emotional utterances from multiple speakers. It was recorded in a controlled anechoic chamber using high-fidelity microphones to ensure that the signals contain no environmental reverberation or background noise. The dataset includes a wide range of emotional states (e.g., neutral, happy, angry, sad) and is sampled at $48$ kHz with clean, full-bandwidth speech for $107$ talkers, varying genders and ages. This dataset is used to sample the source speech signal $S_0(f)$ in the genuine scenario. Measured speech directivities $\mathbf{S}(f,\boldsymbol{\theta}_k)$ are employed from~\cite{Ahrens_TASLP_2021}.

(ii) Gallien \textit{et al.}~\cite{Gallien_AESC_2024} provided four measured loudspeaker directivity impulse responses stored in SOFA (Spatially Oriented Format for Acoustics) files. Each file captures loudspeaker’s directional radiation pattern over a spherical grid using a microphone array placed at multiple azimuths and elevations. The measurements are conducted in anechoic conditions, and each SOFA file includes metadata such as microphone and source positions. This dataset has been employed in the replay scenario to model the directivity of the loudspeaker $\mathbf{L}\!\bigl(f,\boldsymbol{\theta}_k\bigr)$.

\subsection{Background Noise} 

WHAMR!~\cite{Wichern_2019_Interspeech} dataset is used to model background noises $\mathbf{N}(f)$ and $N(f)$ for genuine, spoofing, and replay scenarios in Eqs~\eqref{eq:genuine}--\eqref{eq:replay}. Specifically, mono noise signal from WHAMR!~\cite{Wichern_2019_Interspeech} dataset is rendered directly in multiple channels (e.g., diffuse noise simulated for a microphone array). Instead of duplicating, each channel receives a spatially consistent noise signal using~\cite{mirabilii2021generating}. The noise is then scaled globally to achieve the target \ac{SNR} with respect to the clean multichannel signal.

\begin{table}[h]
\vspace{-6pt}
\caption{Parameters and constraints for data generation.}
\label{tab:sim}
    \centering
        \adjustbox{max width=0.8\textwidth}{%
    \begin{tabular}{cc}
    \hline \hline
    \textbf{Room parameters} & \textbf{Ranges} \\
    \hline
    Room width, height, and length & $\mathcal{U}[3.0, 6.0]$ m \\
    \# of materials (wall, floor, ceiling)  & $13, 7, 8$ \\
    \ac{SNR} & $\mathcal{U}[5, 40]$ dB \\
    \hline
    \textbf{Source parameters} & \textbf{Constraints}\\
    $||\mathbf{p}_{\mathrm{tlk}} - \mathbf{p}_{\mathrm{ASV}} ||$ & $> 1$ m \\
    $||\mathbf{p}_{\mathrm{tlk}} - \mathbf{p}_{\mathrm{spf}} ||$ & $< 1$ m \\
    $||\mathbf{p}_{\mathrm{spk}} - \mathbf{p}_{\mathrm{ASV}} ||$ & $> 1$ m \\    
    Source-to-surface distance & $> 1$ m \\
    Direction of arrival & $\mathcal{U}[-10, 10]$ degrees\\
    \hline \hline
    \end{tabular}
}
\vspace{-7pt}
\end{table}

\subsection{Simulator}
The acoustic simulation framework has been designed using Pyroomacoustics~\cite{Scheibler_ICASSP_2018}, including the generation of spatial \acp{RIR} ($\mathbf{H}_k^{\mathrm{(gen)}}(f)$, $H_k^{\mathrm{(spf)}}(f)$, and $\mathbf{H}_k^{\mathrm{(rep)}}(f)$), room dimensions, and path attenuations.

\begin{figure}[t]
    \centering
    \includegraphics[width=1\linewidth]{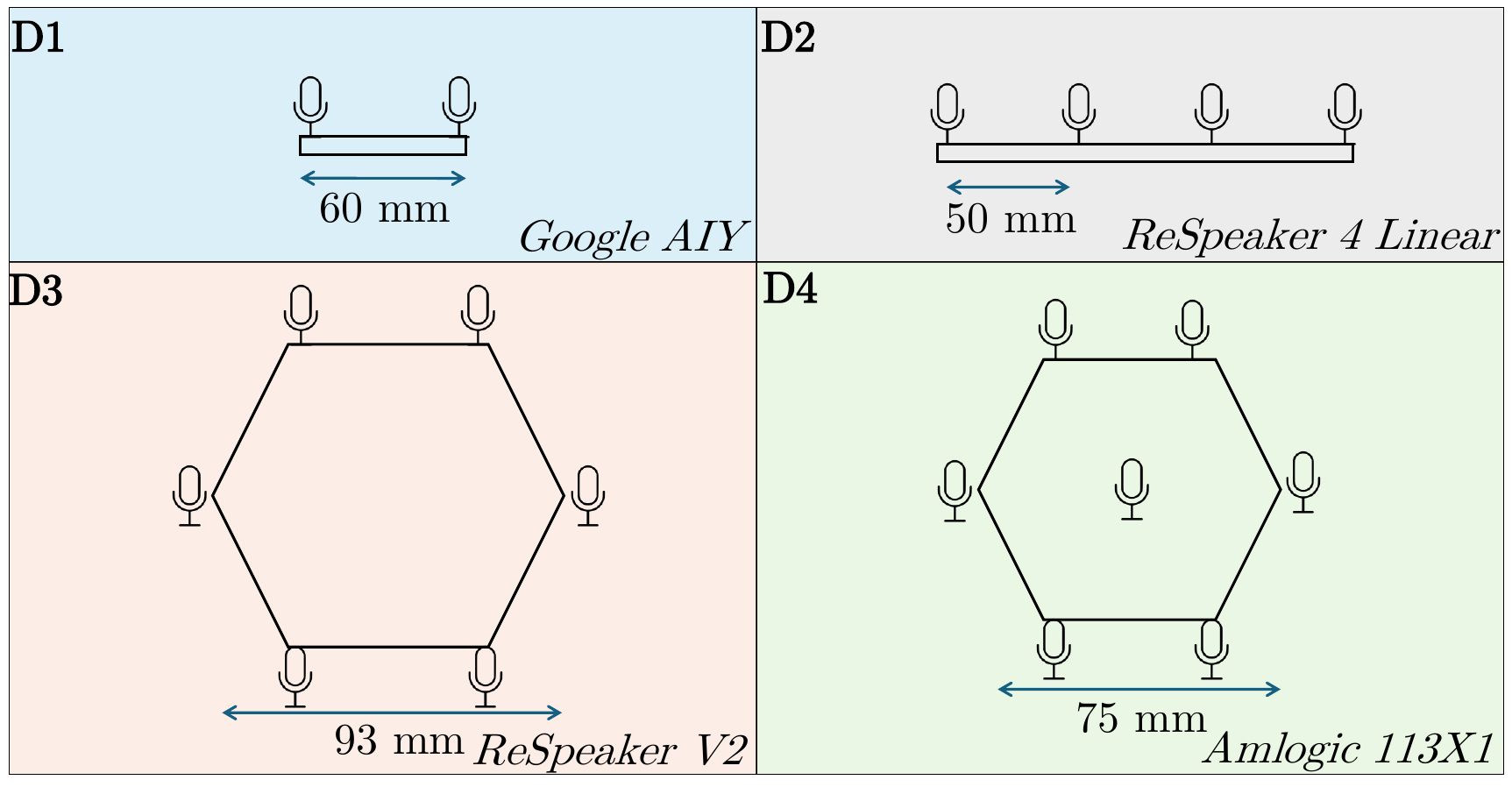}
    \caption{Simulated microphone array with their names, dimensions, and arrangements from ReMASC dataset~\cite{Gong_Interspeech_2019}.}
    \label{fig:micarray}
\end{figure}

We simulate the typical replay attack pipeline, where the attacker first records the speech using a spoofing microphone and then replays the recorded signal through a loudspeaker. The replay signal thus includes both the acoustic coloration introduced during the recording phase and the transformations from the genuine environment. Specifically, $\mathbf{X}_{\mathrm{gen}}(f)$ and $Y_{\mathrm{spf}}(f)$ are collected together in one room. Then, $Y_{\mathrm{spf}}(f)$ is replayed in another room using either a built-in cardioid (frequency-independent) or a SOFA (frequency-dependent) directivities. This design choice is done for avoiding overfitting to specific loudspeaker directivity patterns. By doing so, one genuine and one replay sample are collected for each iteration, forming a balanced dataset.  The array microphone responses $\mathbf{M}_{\mathrm{ASV}}(f, \boldsymbol{\phi}_k^{(\cdot)})$ and the spoofing microphone response $M_{\mathrm{spf}}(f, \boldsymbol{\phi}_k^{(\mathrm{spf})})$ are modeled as omnidirectional with flat frequency response, consistent with the MEMS capsules used in common recording devices. 


The synthetic dataset has been generated by simulating $20$ speeches per talker of $2$ seconds, following~\cite{Gong_Interspeech_2019}, coming from the $107$ talkers of the EARS dataset. The $2$-seconds segments are randomly extracted from the speech corpus. In total, $4,280$ recordings are generated with $2,960$, $400$, and $920$ assigned to training, validation, and testing sets, respectively. The assignment of samples is based on non-overlapping sets of talkers from EARS dataset with a $7$:$1$:$2$ ratio, avoiding any talker leakage information across sets. Parameters for rooms and source-microphone configurations in Pyroomacoustics are detailed in Table~\ref{tab:sim}. In this work we simulated four microphone arrays from ReMASC dataset~\cite{Gong_Interspeech_2019}, which are shown in Figure~\ref{fig:micarray} and briefly described in Section~\ref{sec:remasc}.

\subsection{Real data}\label{sec:remasc}
\ac{ReMASC}~\cite{Gong_Interspeech_2019} has been selected for evaluating the generalization capabilities of trained model on synthetic data since it is the only dataset providing multi-channel recordings for this task. All recordings are synchronized across four parallel \ac{ASV} microphone arrays ${\mathrm{D}1, \mathrm{D}2,\mathrm{D}3,\mathrm{D}4}$ with $2$, $4$, $6$, and $7$ omnidirectional microphones, respectively. Moreover, $\mathrm{D}4$'s sampling rate is $16$ kHz, differently from $\mathrm{D}1$, $\mathrm{D}2$, and $\mathrm{D}3$ which operate at $44.1$ kHz. Bit depth is $16$ bits for all devices, except for $\mathrm{D}3$ which is $32$ bits. Overall, \ac{ReMASC} features $55$ diverse speakers, varying in gender and vocal characteristics, and includes recordings from four distinct environments: an outdoor setting (EnvA), a quiet room with varying source-microphone positions (EnvB), a noisy room with fixed source-microphone positions (EnvC), and a moving vehicle (EnvD), capturing different acoustic conditions. The dataset employs two spoofing microphones to record genuine speech, which is then replayed through four playback devices of varying quality.

\subsection{Feature sets and detection models}

In our experiments, we employ M-ALRAD~\cite{Neri_OJSP_2025} as the multi-channel replay speech detector baseline, a \ac{CRNN}-based approach that jointly processes $C$ multi-channel complex \acp{STFT} $\{X_{{\mathrm{STFT}}_{c,t,f}},\, c = 1, \ldots, C\}$, where $t=1,\ldots,T$ and $f=1,\ldots,F$ index time and frequency bins, respectively. An adaptive complex beamformer $f_{\mathrm{BM}}:\mathbb{C}^{C \times T \times F}  \rightarrow \mathbb{C}^{T \times F}$ first estimates channel-dependent beamforming weights $\mathbf{W} \in \mathbb{C}^{C \times T \times F}$ through a sequence of Conv2D-BatchNorm-ELU-Conv2D operations, where real and imaginary parts are concatenated along the channel dimension. The beamformed spectrogram is then obtained as $\hat{X}_{{\mathrm{STFT}}_{t,f}} = \sum_{c=1}^{C} X_{{\mathrm{STFT}}_{c,t,f}} \cdot w_{c,t,f}$. Three feature maps are extracted from $\hat{X}_{\mathrm{STFT}}$ which are the log-power $\log\lvert\hat{X}_{{\mathrm{STFT}}_{t,f}}\rvert^2$ and the sine and cosine of its phase, following~\cite{Neri_TASLP_2024, Neri_WASPAA_2023}. 

While the beamformer effectively combines spatial information into a single output channel, this aggregation may aggressively collapse and discard inter-channel cues that are diagnostic of replay attacks. In fact, a replay signal undergoes a double acoustic convolution, which alters the inter-channel phase relationships at the microphone array relative to genuine direct-path speech. To preserve these cues, we augment the beamformed feature maps with \ac{IPD} features computed for each pair of adjacent microphones. Given the set of adjacent pairs $\mathcal{P}$ defined by the array geometry, the phase difference for pair $(i,j)\in\mathcal{P}$ at bin $(t,f)$ is $\Delta\phi_{i,j}^{(t,f)} = \angle X_{{\mathrm{STFT}}_{i,t,f}} - \angle X_{{\mathrm{STFT}}_{j,t,f}},$ and sin\&cos are derived from it. 

The resulting $2\lvert\mathcal{P}\rvert$ feature maps are concatenated with the three beamformed feature maps along the channel dimension. The pairs used are adjacent neighbour microphones only: $\lvert\mathcal{P}\rvert = 1$ for ${\mathrm{D}1}$, $3$ for ${\mathrm{D}2}$, and $6$ for the outer ring of ${\mathrm{D}3}$ and ${\mathrm{D}4}$. In addition to phase features, we augment the feature set with per-channel log-power $\log\lvert X_{{\mathrm{STFT}}_{c,t,f}}\rvert^2$, yielding an input of $3 + 2\lvert\mathcal{P}\rvert + C$ channels to the \ac{CRNN} classifier. In this work, we denote this feature as per-ch. 

All models are trained to minimise the binary cross-entropy loss between predicted and true binary labels (genuine or replay). The \ac{STFT} is computed with a Hanning window of length $32$\,ms and $50\,\%$ overlap at $16$\,kHz. Models are trained with a batch size of $16$ using a cosine-annealing scheduled learning rate starting at $0.0003$ for $100$ epochs. Orthogonality and sparsity regularisation losses are applied to the beamforming weights using the same hyperparameters as in~\cite{Neri_OJSP_2025}.

\begin{table*}[t]
\centering
\caption{EER~(\%) on the synthetic test set and four ReMASC environments (5 independent runs, 95\,\%~CI). All models are trained on synthetic data only. \colorbox{light_green}{Green}: EER $<50\%$. \colorbox{light_yellow}{Yellow}: CI overlaps chance. \colorbox{light_red}{Red}: EER $\geq50\%$, CI fully above chance. `—': no recordings available.}
\label{tab:results}
\setlength{\tabcolsep}{4pt}
\begin{adjustbox}{width=0.7\textwidth}

\begin{tabular}{c c c c c c c c}
\toprule
\multirow{2}{*}{\textbf{Model}} & \multirow{2}{*}{\textbf{Dev.}} & \multirow{2}{*}{\textbf{Ch.}} & \multicolumn{5}{c}{\textbf{EER (\%) $\downarrow$}} \\
\cmidrule(lr){4-8}
& & & Synth & EnvA\textsuperscript{(out)} & EnvB\textsuperscript{(ind.1)} & EnvC\textsuperscript{(ind.2)} & EnvD\textsuperscript{(veh)} \\
\midrule
\multirow{4}{*}{M-ALRAD~\cite{Neri_OJSP_2025}} & ${\mathrm{D}1}$ & 2 & \cellcolor{light_green}$14.4{\scriptstyle\,\pm\,1.5}$ & \cellcolor{light_red}$59.9{\scriptstyle\,\pm\,7.8}$ & — & \cellcolor{light_yellow}$37.0{\scriptstyle\,\pm\,13.2}$ & \cellcolor{light_green}$41.9{\scriptstyle\,\pm\,7.6}$ \\
 & ${\mathrm{D}2}$ & 4 & \cellcolor{light_green}$12.0{\scriptstyle\,\pm\,0.4}$ & \cellcolor{light_red}$57.3{\scriptstyle\,\pm\,6.3}$ & \cellcolor{light_yellow}$46.8{\scriptstyle\,\pm\,7.4}$ & \cellcolor{light_yellow}$54.5{\scriptstyle\,\pm\,8.6}$ & \cellcolor{light_yellow}$55.4{\scriptstyle\,\pm\,8.5}$ \\
 & ${\mathrm{D}3}$ & 6 & \cellcolor{light_green}$11.2{\scriptstyle\,\pm\,1.0}$ & \cellcolor{light_yellow}$53.1{\scriptstyle\,\pm\,5.3}$ & \cellcolor{light_yellow}$52.6{\scriptstyle\,\pm\,3.6}$ & \cellcolor{light_green}$36.2{\scriptstyle\,\pm\,7.2}$ & \cellcolor{light_yellow}$51.9{\scriptstyle\,\pm\,9.7}$ \\
 & ${\mathrm{D}4}$ & 7 & \cellcolor{light_green}$10.0{\scriptstyle\,\pm\,1.7}$ & \cellcolor{light_yellow}$48.1{\scriptstyle\,\pm\,13.9}$ & \cellcolor{light_yellow}$56.7{\scriptstyle\,\pm\,8.8}$ & \cellcolor{light_green}$31.8{\scriptstyle\,\pm\,12.3}$ & \cellcolor{light_yellow}$54.3{\scriptstyle\,\pm\,13.5}$ \\
\midrule
\multirow{4}{*}{M-ALRAD~\cite{Neri_OJSP_2025} + \ac{IPD}} & ${\mathrm{D}1}$ & 2 & \cellcolor{light_green}$12.2{\scriptstyle\,\pm\,1.1}$ & \cellcolor{light_red}$58.5{\scriptstyle\,\pm\,4.7}$ & — & \cellcolor{light_green}$28.6{\scriptstyle\,\pm\,2.7}$ & \cellcolor{light_yellow}$48.0{\scriptstyle\,\pm\,6.8}$ \\
 & ${\mathrm{D}2}$ & 4 & \cellcolor{light_green}$9.0{\scriptstyle\,\pm\,1.0}$ & \cellcolor{light_yellow}$47.7{\scriptstyle\,\pm\,6.0}$ & \cellcolor{light_red}$54.8{\scriptstyle\,\pm\,3.3}$ & \cellcolor{light_green}$41.6{\scriptstyle\,\pm\,6.5}$ & \cellcolor{light_yellow}$50.0{\scriptstyle\,\pm\,9.6}$ \\
 & ${\mathrm{D}3}$ & 6 & \cellcolor{light_green}$8.9{\scriptstyle\,\pm\,1.4}$ & \cellcolor{light_green}$39.9{\scriptstyle\,\pm\,6.2}$ & \cellcolor{light_red}$61.8{\scriptstyle\,\pm\,7.4}$ & \cellcolor{light_green}$19.2{\scriptstyle\,\pm\,8.9}$ & \cellcolor{light_yellow}$56.7{\scriptstyle\,\pm\,15.8}$ \\
 & ${\mathrm{D}4}$ & 7 & \cellcolor{light_green}$8.6{\scriptstyle\,\pm\,1.3}$ & \cellcolor{light_yellow}$38.1{\scriptstyle\,\pm\,16.9}$ & \cellcolor{light_red}$60.3{\scriptstyle\,\pm\,4.9}$ & \cellcolor{light_green}$41.8{\scriptstyle\,\pm\,6.8}$ & \cellcolor{light_green}$41.7{\scriptstyle\,\pm\,6.2}$ \\
\midrule
\multirow{4}{*}{M-ALRAD~\cite{Neri_OJSP_2025} + per-ch + IPD} & ${\mathrm{D}1}$ & 2 & \cellcolor{light_green}$11.9{\scriptstyle\,\pm\,0.6}$ & \cellcolor{light_red}$57.8{\scriptstyle\,\pm\,7.1}$ & — & \cellcolor{light_green}$27.9{\scriptstyle\,\pm\,3.6}$ & \cellcolor{light_yellow}$47.2{\scriptstyle\,\pm\,5.0}$ \\
 & ${\mathrm{D}2}$ & 4 & \cellcolor{light_green}$9.3{\scriptstyle\,\pm\,1.1}$ & \cellcolor{light_red}$57.7{\scriptstyle\,\pm\,3.3}$ & \cellcolor{light_red}$58.4{\scriptstyle\,\pm\,2.9}$ & \cellcolor{light_yellow}$45.3{\scriptstyle\,\pm\,9.5}$ & \cellcolor{light_yellow}$45.9{\scriptstyle\,\pm\,5.6}$ \\
 & ${\mathrm{D}3}$ & 6 & \cellcolor{light_green}$8.3{\scriptstyle\,\pm\,0.7}$ & \cellcolor{light_green}$43.5{\scriptstyle\,\pm\,4.5}$ & \cellcolor{light_red}$61.5{\scriptstyle\,\pm\,5.4}$ & \cellcolor{light_green}$23.6{\scriptstyle\,\pm\,5.6}$ & \cellcolor{light_yellow}$51.8{\scriptstyle\,\pm\,10.1}$ \\
 & ${\mathrm{D}4}$ & 7 & \cellcolor{light_green}$8.7{\scriptstyle\,\pm\,1.4}$ & \cellcolor{light_green}$37.0{\scriptstyle\,\pm\,12.0}$ & \cellcolor{light_red}$60.7{\scriptstyle\,\pm\,4.1}$ & \cellcolor{light_green}$29.3{\scriptstyle\,\pm\,6.0}$ & \cellcolor{light_yellow}$48.7{\scriptstyle\,\pm\,6.7}$ \\
\bottomrule
\end{tabular}
\end{adjustbox}
\end{table*}

\section{Evaluation}
\label{sec:results}
In this section we train and evaluate the models on the simulated replay speech dataset and ReMASC. An analysis on the selected features for the replay speech detection and generalization to unseen environments is carried out. To assess the performance of the detection, we employ the \ac{EER} metric, following previous works~\cite{zhang2016voicelive, Sanchez_TIFS_2026, Neri_OJSP_2025}. All the results from $5$ independent trainings have been collected to compute the $95\%$ confidence interval.

\subsection{Results}
Table~\ref{tab:results} reports the \ac{EER} of the baseline M-ALRAD~\cite{Neri_OJSP_2025} and the two feature extensions across the four ReMASC devices and environments, all trained exclusively on synthetic data. On the synthetic test set, both proposed models consistently outperform the baseline across all devices, with \ac{EER} reductions ranging from $1.3$ to $2.2$ percentage points for M-ALRAD + \ac{IPD} and from $0.7$ to $2.5$ percentage points for M-ALRAD + per-ch + \ac{IPD}, confirming that the additional feature branches provide discriminative spatial information beyond what the beamformer captures. Furthermore, the synthetic \ac{EER} decreases monotonically with the number of microphone channels for both proposed models, whereas the baseline does not exhibit this trend consistently, suggesting that \ac{IPD} features are necessary for the model to exploit the spatial diversity provided by larger arrays.

As demonstrated from previous work~\cite{Sanchez_TIFS_2026} with similar performance, generalisation to real recordings is challenging and strongly environment-dependent. EnvA (outdoor) benefits substantially from the proposed models for ${\mathrm{D}3}$ and ${\mathrm{D}4}$, with M-ALRAD + \ac{IPD} achieving $39.9 \pm 6.2\%$ and $38.1 \pm 16.9\%$ respectively, compared to $53.1 \pm 5.3\%$ and $48.1 \pm 13.9\%$ for the baseline. The outdoor improvement suggests that \ac{IPD} features are slightly transferable to uncontrolled acoustic conditions, as propagation delays depend primarily on geometry rather than on room-specific spectral colouration. 

EnvB (indoor quiet study room) remains challenging for all models and all devices, with \ac{EER} consistently at or above chance. Since this behaviour is invariant to the feature and microphone configuration, it cannot be attributed to the choice of feature branch or to model design, and instead reflects a mismatch in the input distribution between the synthetic and real data. The most likely cause is the source–array geometry: our training set confines the talker to a near-frontal cone (withing $10$ degrees), whereas EnvB is recorded with widely varying source–microphone positions, so its arrival directions are largely out-of-distribution for the geometry-dependent \ac{IPD} and beamformer features. A background-noise $\mathbf{N}(f)$ mismatch may also contribute; the two factors are confounded, since EnvC is fixed-position and noisy while EnvB is varying-position and quiet, and they cannot be separated with the present data. 

EnvC (indoor lounge) yields the best results, with both proposed features configurations achieving \ac{EER} below $50\%$ across all devices (down to $19.2 \pm 8.9\%$ for M-ALRAD + \ac{IPD} with ${\mathrm{D}3}$, and $23.6 \pm 5.6\%$ for M-ALRAD + per-ch + \ac{IPD} with ${\mathrm{D}3}$, compared to $36.2 \pm 7.2\%$ for the baseline). EnvD (moving vehicle) also shows small improvements for the larger arrays, with ${\mathrm{D}4}$ reaching $41.7 \pm 6.2\%$ with M-ALRAD + \ac{IPD}. However, as in EnvA, the performance is expected to be scarse due to the challenging environment that is not modeled by the simulator. Overall, the feature modifications to M-ALRAD perform comparably better than the baseline, with M-ALRAD + \ac{IPD} achieves better \ac{EER} on EnvA and EnvD for most devices, while M-ALRAD + per-ch + \ac{IPD} yields slightly lower \ac{EER} on EnvC for ${\mathrm{D}3}$ and ${\mathrm{D}4}$. The difference between the variants, and hence the different contribution of phase and magnitude features, is not consistent across environments, motivating the ablation study in Section~\ref{sec:ablation}.

\begin{table*}[t]
\centering
\caption{\ac{EER}~(\%) averaged across the four microphone arrays ($20$~runs per variant, $95$\,\%~CI). BF = adaptive complex beamformer; per-ch = per-channel log-power; IPD = inter-channel phase differences. \colorbox{light_green}{Green}: EER $<50\%$. \colorbox{light_yellow}{Yellow}: CI overlaps chance. \colorbox{light_red}{Red}: \ac{EER} $\geq50\%$, CI fully above chance.}
\label{tab:ablation}
\setlength{\tabcolsep}{4pt}

\begin{adjustbox}{width=0.8\textwidth}
\begin{tabular}{c c c c c c c c c}
\toprule
\multirow{2}{*}{\textbf{Features}} & \multirow{2}{*}{\textbf{BF}} & \multirow{2}{*}{\textbf{per-ch}} & \multirow{2}{*}{\textbf{IPD}} & \multicolumn{5}{c}{\textbf{EER (\%) $\downarrow$}} \\
\cmidrule(lr){5-9}
 & & & & Synth & EnvA\textsuperscript{(out)} & EnvB\textsuperscript{(ind.1)} & EnvC\textsuperscript{(ind.2)} & EnvD\textsuperscript{(veh)} \\
\midrule
BF & \checkmark & — & — & \cellcolor{light_green}$12.2{\scriptstyle\,\pm\,1.0}$ & \cellcolor{light_red}$53.6{\scriptstyle\,\pm\,3.1}$ & \cellcolor{light_yellow}$53.6{\scriptstyle\,\pm\,6.7}$ & \cellcolor{light_green}$35.6{\scriptstyle\,\pm\,5.3}$ & \cellcolor{light_yellow}$49.3{\scriptstyle\,\pm\,4.1}$ \\
BF + per-ch & \checkmark & \checkmark & — & \cellcolor{light_green}$11.7{\scriptstyle\,\pm\,0.7}$ & \cellcolor{light_red}$55.2{\scriptstyle\,\pm\,4.3}$ & \cellcolor{light_red}$58.0{\scriptstyle\,\pm\,6.3}$ & \cellcolor{light_green}$36.2{\scriptstyle\,\pm\,4.2}$ & \cellcolor{light_yellow}$51.6{\scriptstyle\,\pm\,4.6}$ \\
BF + IPD & \checkmark & — & \checkmark & \cellcolor{light_green}$9.7{\scriptstyle\,\pm\,0.8}$ & \cellcolor{light_yellow}$46.1{\scriptstyle\,\pm\,5.1}$ & \cellcolor{light_red}$59.0{\scriptstyle\,\pm\,2.8}$ & \cellcolor{light_green}$33.3{\scriptstyle\,\pm\,5.3}$ & \cellcolor{light_yellow}$49.1{\scriptstyle\,\pm\,4.4}$ \\
BF + per-ch + IPD & \checkmark & \checkmark & \checkmark & \cellcolor{light_green}$\mathbf{9.5{\scriptstyle\,\pm\,0.8}}$ & \cellcolor{light_yellow}$49.2{\scriptstyle\,\pm\,4.9}$ & \cellcolor{light_red}$60.2{\scriptstyle\,\pm\,1.9}$ & \cellcolor{light_green}$\mathbf{31.5{\scriptstyle\,\pm\,4.6}}$ & \cellcolor{light_yellow}$48.4{\scriptstyle\,\pm\,2.7}$ \\
\midrule
per-ch & — & \checkmark & — & \cellcolor{light_green}$17.3{\scriptstyle\,\pm\,1.3}$ & \cellcolor{light_red}$57.4{\scriptstyle\,\pm\,5.1}$ & \cellcolor{light_yellow}$52.3{\scriptstyle\,\pm\,4.3}$ & \cellcolor{light_yellow}$50.8{\scriptstyle\,\pm\,5.9}$ & \cellcolor{light_yellow}$52.8{\scriptstyle\,\pm\,3.0}$ \\
IPD & — & — & \checkmark & \cellcolor{light_green}$11.3{\scriptstyle\,\pm\,1.5}$ & \cellcolor{light_yellow}$51.8{\scriptstyle\,\pm\,5.5}$ & \cellcolor{light_yellow}$52.5{\scriptstyle\,\pm\,4.5}$ & \cellcolor{light_green}$38.9{\scriptstyle\,\pm\,3.8}$ & \cellcolor{light_yellow}$53.1{\scriptstyle\,\pm\,4.2}$ \\
per-ch + IPD & — & \checkmark & \checkmark & \cellcolor{light_green}$10.1{\scriptstyle\,\pm\,1.0}$ & \cellcolor{light_yellow}$51.1{\scriptstyle\,\pm\,4.4}$ & \cellcolor{light_red}$59.6{\scriptstyle\,\pm\,5.5}$ & \cellcolor{light_green}$35.0{\scriptstyle\,\pm\,5.2}$ & \cellcolor{light_yellow}$51.3{\scriptstyle\,\pm\,3.0}$ \\
\bottomrule
\end{tabular}
\end{adjustbox}

\end{table*}

\subsection{Ablation Study}
\label{sec:ablation}

Table~\ref{tab:ablation} reports the results of the ablation across the three feature branches, averaged over the four microphone arrays. It is worth noting that the ablation study removes features from the first layer of the \ac{CRNN} classifier to analyze the impact of each feature to the detection task.

\textbf{Role of the beamformer.} Removing the beamformer while retaining only per-channel log-power features (per-ch, no BF) yields a synthetic \ac{EER} of  $17.3 \pm 1.3\%$, which is $5.1$ percentage points worse than the BF-only baseline ($12.2 \pm 1.0\%$), and near-chance performance on all real environments. This confirms that per-channel amplitude features alone cannot learn the replay detection task without the spatial coherence provided by the adaptive beamformer, which effectively acts as a learned spatial pre-filter. IPD features without the beamformer (IPD, no BF) perform more competitively on the synthetic task ($11.3 \pm 1.5\%$) and achieve $38.9 \pm 3.8\%$ on EnvC, but remain worse than the BF + IPD configuration ($9.7 \pm 0.8\%$ synthetic, $33.3 \pm 5.3\%$ EnvC), demonstrating that the beamformer and \ac{IPD} features are complementary rather than redundant.

\textbf{Role of IPD features.} Adding \ac{IPD} to the beamformer (BF + IPD) reduces synthetic \ac{EER} by $2.5$ percentage points relative to BF alone and improves EnvA by $7.5$ percentage points, that is one of the largest single-branch gain in this work. This contrasts sharply with adding per-channel log-power to the beamformer (BF + per-ch), which yields a marginal synthetic improvement of $0.5$ percentage points and degrades the performance on EnvA by $1.6$ and on EnvB by $4.4$ percentage points. The degradation indicates that per-channel amplitude features learned from simulated noise and room materials do not transfer reliably to real recordings, whereas inter-channel phase differences, which depend on propagation geometry rather than spectral texture, generalise more consistently.

\textbf{EnvB.} All variants, including those without the beamformer, produce \ac{EER} at or above chance on EnvB. Since this behaviour is invariant to the feature configuration, it cannot be attributed to model design choices and is instead consistent with a combination of background noise $\mathbf{N}(f)$ and varying microphone-source positions being modeled differently in the simulator with respect to real recordings.

\begin{figure}[h]
    \centering
    \includegraphics[width=\linewidth]{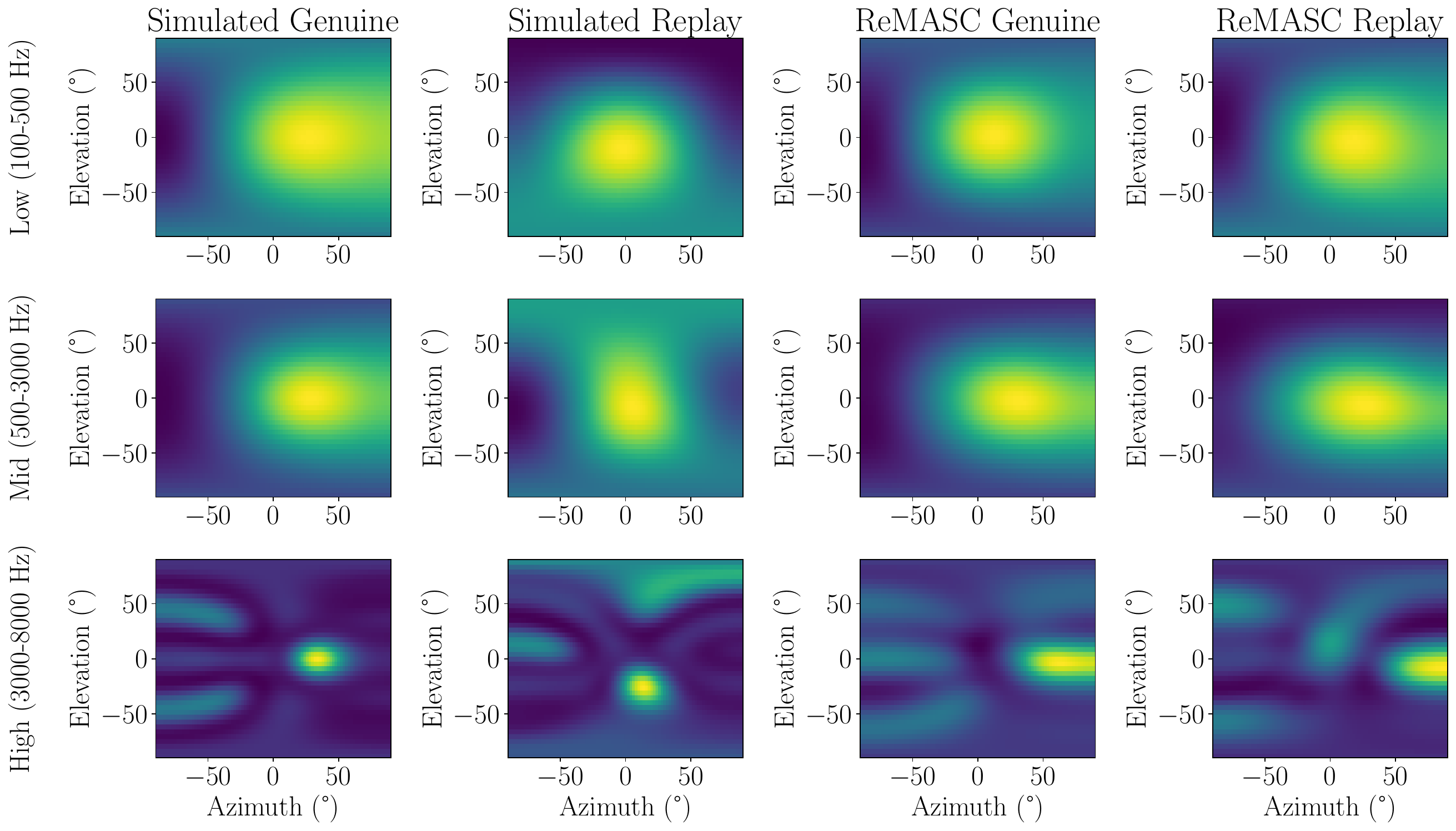}
    \caption{Average delay-and-sum acoustic maps for device $\mathrm{D}3$, comparing simulated recordings against real ReMASC EnvC\textsuperscript{(ind.2)} recordings for the genuine and replay conditions. Rows correspond to the low ($100$–$500$ Hz), mid ($500$–$3000$ Hz), and high ($3000$–$8000$ Hz) bands, following the same parameters as~\cite{Neri_EUSIPCO_2026}; each panel is the steered-response power averaged over all recordings of the corresponding condition and min-max normalised independently to $[0, 1]$. Axes denote the azimuth and elevation of the look direction.}
    \label{fig:AM}
\end{figure}

\subsection{Real-Synthetic Comparison}
To compare the simulated recordings with real data, we employ high-level features as acoustic maps~\cite{Neri_EUSIPCO_2026}. Fig.~\ref{fig:AM} shows average delay-and-sum acoustic maps over the steered azimuth–elevation grid for device $\mathrm{D}3$ and compare the simulated genuine and replay conditions against their real EnvC counterparts. Since each panel is normalised independently, the comparison reflects the distribution and shape of the steered-response power rather than absolute levels. In the low and mid bands (first two rows) both the simulated and the real maps concentrate energy in a single dominant lobe close to the array broadside, with comparable angular spread for both the genuine and the replay conditions. This indicates that the simulator reproduces the direct-path arrival direction and the overall spatial energy distribution of the EnvC source. The similarity between the genuine and replay maps, in both the simulated and the real data, is expected given the near-frontal source placement in this environment. In the high band, both real and simulated maps exhibit the symmetric sidelobe pattern characteristic of spatial aliasing for this array (as inter-element spacing is around $46.5$ mm, with aliasing onset near $3.7$ kHz), so the peak location is less reliable there. Despite of this, the qualitative structure is consistent across simulation and reality. The residual differences are consistent with the diffuse noise model and the absence of explicitly localised interferers in the simulation.

\section{Conclusion}
\label{sec:conclusion}
In this work, we proposed an acoustic simulation framework for multi-channel replay speech detection and used it to train and evaluate spatial replay detectors on the ReMASC corpus without access to any real training data. We extended M-ALRAD with \ac{IPD} features that preserve directional cues discarded by the beamformer's channel collapse, and carried out a full ablation study to isolate the contribution of each feature branch. The ablation confirmed that the adaptive beamformer and \ac{IPD} features are complementary and both necessary: removing the beamformer while retaining \ac{IPD} consistently degrades performance, whereas adding \ac{IPD} to the beamformer provides the largest single generalisation gain in the study. Per-channel log-power features, by contrast, learn simulation-specific spectral textures that do not transfer reliably to real recordings. EnvB (indoor quiet room) remains an open challenge for all configurations. Rather than a model-design limitation, this reflects a distribution mismatch in the input data that is most plausibly the restricted range of source–array geometries in our training set relative to EnvB's widely varying positions. Future work will broaden the simulated source–array geometries and noise conditions, and incorporate a wider variety of measured loudspeaker and room configurations, to improve generalisation across unseen environments.

 
\bibliographystyle{IEEEtran}
\bibliography{ref}

\end{document}